\documentclass[prl,twocolumn,showpacs,amsmath,amssymb]{revtex4-1}

\usepackage{subfigure}
\usepackage{graphicx}
\usepackage{epstopdf}
\usepackage{soul}
\usepackage[mathlines]{lineno}
\usepackage{dcolumn}
\usepackage{color}
\usepackage{bm}
\usepackage{overpic}
\usepackage{rotating}
\usepackage{times}
\usepackage{multirow}
\usepackage{float}
\usepackage{mathrsfs}
\usepackage{booktabs}
\usepackage{float}
\usepackage[mathlines]{lineno}
\usepackage[T1]{fontenc}
\usepackage{xcolor}

\newcommand{\BR}{{\cal B}}

\newcommand{\pip}{\pi^+}
\newcommand{\pim}{\pi^-}

\newcommand{\etap}{\eta^\prime}

\newcommand{\jpsi}{J/\psi}

\newcommand{\etamp}{\eta_{1}(1855)}

\newcommand{\pimp}{\pi^+\pi^-}

\begin{document}
\lefthyphenmin=2
\righthyphenmin=2
\uchyph=0
\tolerance=1000

\normalsize
\parskip=5pt plus 1pt minus 1pt

\title{\boldmath Observation of an Isoscalar Resonance with Exotic $J^{PC}=1^{-+}$ Quantum Numbers in $\jpsi\rightarrow\gamma\eta\eta'$}
\author{
M.~Ablikim$^{1}$, M.~N.~Achasov$^{10,b}$, P.~Adlarson$^{68}$, S. ~Ahmed$^{14}$, M.~Albrecht$^{4}$, R.~Aliberti$^{28}$, A.~Amoroso$^{67A,67C}$, M.~R.~An$^{32}$, Q.~An$^{64,50}$, X.~H.~Bai$^{58}$
% ...
    \\
        \vspace{0.2cm}
        (BESIII Collaboration)\\
        \vspace{0.2cm} {\it
$^{1}$ Institute of High Energy Physics, Beijing 100049, People's Republic of China\\
% ...
$^{4}$ Bochum Ruhr-University, D-44780 Bochum, Germany\\
% ...
$^{10}$ G.I. Budker Institute of Nuclear Physics SB RAS (BINP), Novosibirsk 630090, Russia\\
% ...
$^{14}$ Helmholtz Institute Mainz, Staudinger Weg 18, D-55099 Mainz, Germany\\
% ...
$^{28}$ Johannes Gutenberg University of Mainz, Johann-Joachim-Becher-Weg 45, D-55099 Mainz, Germany\\
% ...
$^{32}$ Liaoning Normal University, Dalian 116029, People's Republic of China\\
% ...
$^{50}$ State Key Laboratory of Particle Detection and Electronics, Beijing 100049, Hefei 230026, People's Republic of China\\
% ...
$^{58}$ University of Jinan, Jinan 250022, People's Republic of China\\
% ...
$^{64}$ University of Science and Technology of China, Hefei 230026, People's Republic of China\\
% ...
$^{67}$ University of Turin and INFN, (A)University of Turin, I-10125, Turin, Italy; (B)University of Eastern Piedmont, I-15121, Alessandria, Italy; (C)INFN, I-10125, Turin, Italy\\
$^{68}$ Uppsala University, Box 516, SE-75120 Uppsala, Sweden\\
% ...
$^{b}$ Also at the Novosibirsk State University, Novosibirsk, 630090, Russia\\
% ...
        }
}

\date{\today}

\begin{abstract} 
Using a sample of (10.09$\pm$0.04)$\times$10$^{9}$ $\jpsi$ events collected with the BESIII detector operating at the BEPCII storage ring, a partial wave analysis of the decay $\jpsi \rightarrow \gamma\eta\etap$ is performed. The first observation of an isoscalar state with exotic quantum numbers $J^{PC}=1^{-+}$, denoted as $\etamp$, is reported in the process $\jpsi \rightarrow \gamma\etamp$ with $\etamp\rightarrow\eta\etap$. 
Its mass and width are measured to be (1855$\pm$9$_{-1}^{+6}$)~MeV/$c^{2}$ and (188$\pm$18$_{-8}^{+3}$)~MeV, respectively, where the first uncertainties are statistical and the second are systematic, and its statistical significance is estimated to be larger than 19$\sigma$.

\end{abstract}

\maketitle

The quark model describes a conventional meson as a bound state of  a quark and an antiquark. However, due to the non-Abelian nature of QCD, bound states with gluonic degrees of freedom, such as glueballs  and hybrids, are also expected.  
The clear identification of these QCD exotics would validate and advance our quantitative understanding of QCD.
Radiative decays of the $\jpsi$ meson provide a gluon-rich environment and are therefore regarded as one of the most promising hunting grounds for gluonic exciations~\cite{Cakir:1994jf,Close:1996yc,Sarantsev:2021ein,Rodas:2021tyb}. 

Hybrid mesons are $q\overline q$ states with explicit excitations of the gluon field.
They were first proposed several decades 
ago~\cite{Horn:1977rq,Isgur:1984bm,Chanowitz:1982qj,Barnes:1982tx,Close:1994hc}, 
and have been the source of more recent
lattice QCD (LQCD)~\cite{Lacock:1996ny,MILC:1997usn,Dudek:2011bn,Dudek:2013yja} and phenomenological QCD studies~\cite{Szczepaniak:2001rg,Szczepaniak:2006nx,Guo:2008yz,Bass:2018uon}. 
Models and LQCD predict that the exotic $J^{PC} = 1^{-+}$ nonet of hybrid mesons 
is the lightest, 
with a mass around 1.7-2.1~GeV/$c^{2}$~\cite{Meyer:2015eta,Dudek:2013yja,Lacock:1996ny}. 
The predicted decay widths are model dependent; most hybrids are expected to be rather broad, but some can be as narrow as 100 MeV~\cite{Page:1998gz}. 
There are currently three $1^{-+}$ candidates: the $\pi_{1}(1400)$, $\pi_{1}(1600)$, and $\pi_{1}(2015)$~\cite{Meyer:2010ku, Klempt:2007cp,JPAC:2018zyd,Woss:2020ayi}, which are all isovector states.
Finding an isoscalar $1^{-+}$ hybrid state is critical for establishing the hybrid multiplet. Decaying to $\eta\eta'$ in a $P$ wave is expected for an isoscalar $1^{-+}$ hybrid state~\cite{bib_etaetap_Pwave_1,bib_etaetap_Pwave_2,Eshraim:2020ucw}.


In this Letter, a partial wave analysis (PWA) of the process $\jpsi\rightarrow\gamma\eta\etap$ is performed.
The first observation of an isoscalar state with exotic quantum numbers $J^{PC}= 1^{-+}$, denoted as $\etamp$,
is reported with high statistical significance in the decay chain 
$\jpsi \rightarrow \gamma\etamp \rightarrow \gamma\eta\etap$.
In addition, a large $\jpsi\rightarrow\gamma f_{0}(1500)\rightarrow\gamma\eta\etap$ component is observed, while $\jpsi\rightarrow\gamma f_{0}(1710)\rightarrow\gamma\eta\etap$ is found to be insignificant. 
More details are presented in a companion paper~\cite{PRD}. The analysis is based on (10.09$\pm$0.04)$\times$10$^{9}$ $\jpsi$ events accumulated with the BESIII detector~\cite{Ablikim:NumOfJpsi} operating at the BEPCII storage ring. A detailed description of the BESIII detector can be found in Ref.~\cite{Ablikim:2009aa}.


Candidate events for the process $\jpsi\rightarrow\gamma\eta\etap$ are selected using the criteria described in Ref.~\cite{PRD}.
The $\eta$ is reconstructed via the decay channel $\gamma\gamma$ and the $\eta'$ is reconstructed via  
$\eta'\rightarrow\gamma\pimp$ and $\eta'\rightarrow\eta\pimp$. 
Backgrounds are estimated by the $\etap$ mass sidebands, with details given in Ref.~\cite{PRD}. 
For $\jpsi\rightarrow\gamma\eta\etap$, $\etap\rightarrow\eta\pimp$, the selected sample contains a total of 4788 candidate events including 391$\pm9$ background events, while for $\jpsi\rightarrow\gamma\eta\etap$, $\etap\rightarrow\gamma\pimp$, 
there are 10\,544 total events including 1336$\pm21$ background events.


Using the GPUPWA framework~\cite{gpuframework}, 
a PWA is performed  for
the selected candidate events from the process $\jpsi$ $\rightarrow$ $\gamma\eta\etap$ with $\etap$ $\rightarrow$ $\gamma\pimp$ and $\etap$ $\rightarrow$ $\eta\pimp$.
Quasi-two-body decay amplitudes in the sequential decay processes $\jpsi \rightarrow \gamma X, X\rightarrow\eta\etap$ and $\jpsi \rightarrow\eta X, X\rightarrow \gamma\etap$ and $\jpsi \rightarrow\etap X, X\rightarrow \gamma\eta$ are constructed using the covariant tensor amplitudes described in Ref.~\cite{bib19}.  
The resonance $X$ is parametrized by a relativistic Breit-Wigner (BW) propagator with constant width. 
The complex coefficients of the amplitudes (relative magnitudes and phases) and resonance parameters (masses and widths) are determined by an unbinned maximum likelihood fit to the data. 
The joint probability  for observing the $N$ events in the data sample is 
\begin{eqnarray}\label{joint probability density}
\mathcal{L} \equiv \prod\limits_{i=1}^{N}\frac{ \left|M(\xi_{i}) \right|^2\epsilon(\xi_{i})\Phi(\xi_{i}) }{\sigma^\prime},
\end{eqnarray}
where $\epsilon(\xi_{i})$ is the detection efficiency,
 $\Phi(\xi_{i})$ is the standard element of phase space, and $M(\xi_{i})= \sum_{X} A_X(\xi_i)$ is the matrix element describing the decay processes from the $J/\psi$ to the final state $\gamma\eta\eta^\prime$.  $A_X(\xi_i)$ is the amplitude corresponding to intermediate resonance $X$. Details of the likelihood function construction can be found in Ref.~\cite{PRD}. The free parameters are optimized using MINUIT~\cite{minuit}. 
To account for background,
the background contribution to the likelihood function is estimated using $\eta'$ sideband events and is subtracted from the total log-likelihood value~\cite{Langenbruch:2019nwe}. 
The two decay channels $\jpsi \rightarrow \gamma\eta\etap, \etap \rightarrow \eta \pip \pim$ and $\jpsi \rightarrow \gamma\eta\etap, \etap \rightarrow \gamma \pip \pim$ are combined by adding their log-likelihood values;
they share the same set of masses, widths, relative magnitudes, and phases. 

The set of two-body amplitudes used in the PWA is determined in three steps.
First, a "PDG-optimized" set of amplitudes is determined.  
To describe the $\eta\etap$ spectrum, all kinematically allowed resonances with $J^{PC} =$ $0^{++}$, $2^{++}$, and $4^{++}$ listed in the PDG~\cite{Zyla:2020zbs}, Ref.~\cite{Bugg:2004xu}, and Ref.~\cite{BESIII:2012rtd} are considered.
Similarly, to describe the $\gamma\eta^{(\prime)}$ spectrum, all resonances listed in the PDG with $J^{PC} =$ $1^{+-}$ and  $1^{--}$ are considered. 
All possible combinations of these resonances are evaluated. 
The statistical significance for each resonance is determined by examining the probability of the change in log-likelihood values when including and excluding this resonance in the fits, 
where the probability is calculated under the $\chi^{2}$ distribution hypothesis taking into account the change in the number of degrees of freedom.
The masses and widths of the resonances near $\eta\eta'$ threshold [$f_{0}(1500)$, $f_{2}(1525)$, $f_{2}(1565)$, and $f_{2}(1640)$] 
as well as those with small fit fractions~($\textless$3$\%$)
are always fixed to the PDG~\cite{Zyla:2020zbs} values. 
The mass and width of the $f_{0}(2330)$, which corresponds to a clear structure around 2.3 GeV$c^{2}$ in the $\eta\eta'$ mass spectrum, are free parameters. All other masses and widths are also free parameters in the fit.
The final PDG-optimized set of amplitudes is the combination where each included resonance has a statistical significance larger than 5$\sigma$. 

In the second step,
a search is performed for additional resonances with $J^{PC}=1^{-+}, 0^{++}, 2^{++}, 4^{++}, 1^{+-}_{\gamma\eta^{(\prime)}}$, and $1^{--}_{\gamma\eta^{(\prime)}}$ by individually adding each possibility to the PDG-optimized solution and scanning  over its mass and width.
The significance of each additional resonance at each mass and width is evaluated.
The result indicates that a significant $1^{-+}$ contribution ($\textgreater$$7\sigma$) is needed around 1.9 GeV in the $\eta\etap$ system. The significances for all other additional contributions are less than 5$\sigma$.
Therefore, an $\eta_1$ state is included in the PWA.

In the third step, a baseline set of amplitudes is determined that includes the $\eta_1$ state with 
its mass and width as free parameters. 
The statistical significances of all resonances in the PDG-optimized set are reevaluated in the presence of the $\eta_1$ state.  Resonances with significance less than $5\sigma$ are removed.  
The resulting baseline set of amplitudes contains a significant 
contribution from an isoscalar state with exotic quantum numbers $J^{PC} = 1^{-+}$, denoted as $\etamp$. Its statistical significance is 21.4$\sigma$, and its mass and width are (1855$\pm$9$_{\rm {stat}}$)~MeV/$c^{2}$ and (188$\pm$18$_{\rm {stat}}$)~MeV, respectively.
In addition, the baseline set of amplitudes includes 
four $0^{++}$ resonances [$f_{0}(1500)$, $f_{0}(1810)$, $f_{0}(2020)$, $f_{0}(2330)$], two $2^{++}$ resonances [$f_{2}(1565)$, $f_{2}(2010)$], a nonresonant contribution modeled by a $0^{++}$ $\eta\eta^{\prime}$ system uniformly distributed in phase space (PHSP), and two $1^{+-}$ resonances [$h_{1}(1415)$, $h_{1}(1595)$] in the $\gamma\eta$ system.
In addition, a $4^{++}$ resonance $f_{4}(2050)$ with statistical significance 4.6$\sigma$ is included.

The results of the PWA with the baseline set of amplitudes, including 
the masses and widths of the resonances, the product branching fractions $\jpsi\rightarrow\gamma X\rightarrow\gamma\eta\eta'$ or $\jpsi\rightarrow\eta^{(\prime)} X\rightarrow\gamma\eta\eta'$, and the statistical significances, are summarized in Table~\ref{Summary of all}.
The measured masses and widths of the $f_{0}(2020)$ and $f_{2}(2010)$ are consistent with the PDG~\cite{Zyla:2020zbs} average values. 
The measured mass of the $f_{0}(2330)$, which is unestablished in the PDG~\cite{Zyla:2020zbs}, is consistent with the results of Ref.~\cite{Bugg:2004xu}, but our measured width is 79 MeV smaller (3.4$\sigma$).

\begin{figure*}[htbp]
  \centering

    \subfigure{  	
    \includegraphics[width=0.34\textwidth]{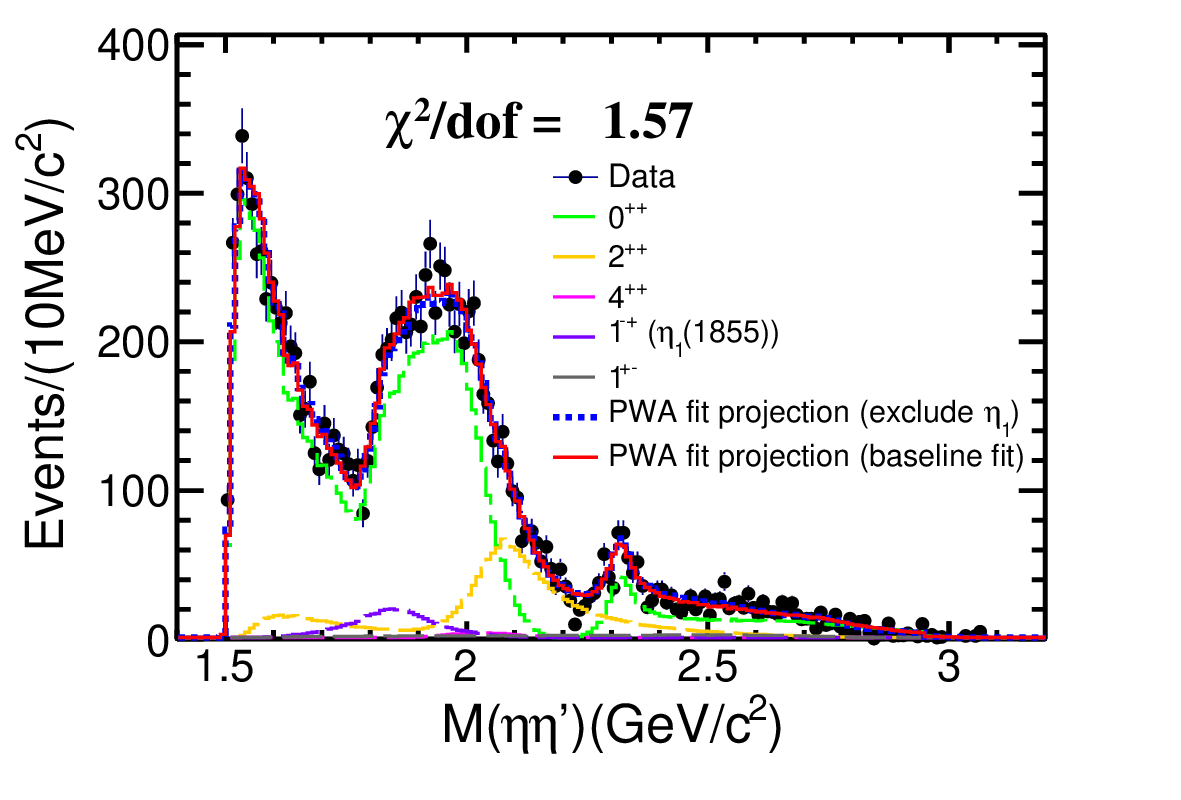}\put(-129,97){(a)}}
  \subfigure{
    \includegraphics[width=0.34\textwidth]{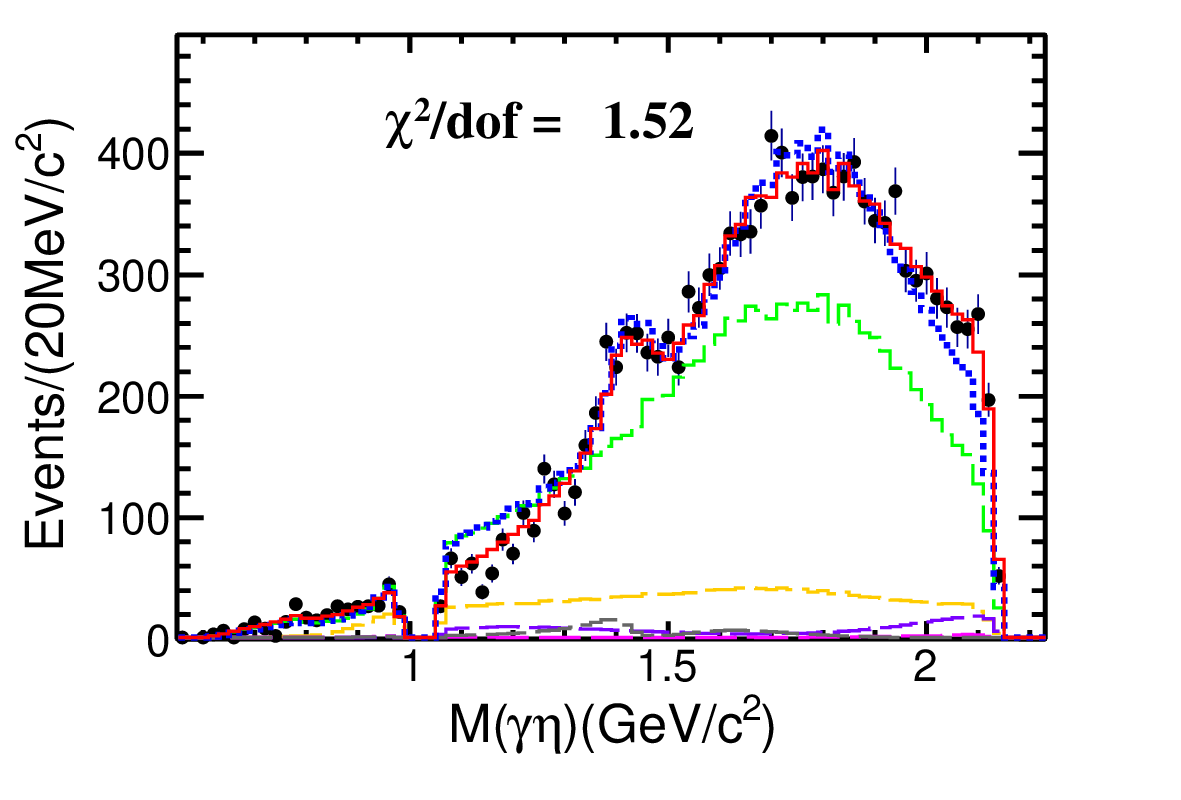}\put(-129,97){(b)}}
  \subfigure{
    \includegraphics[width=0.34\textwidth]{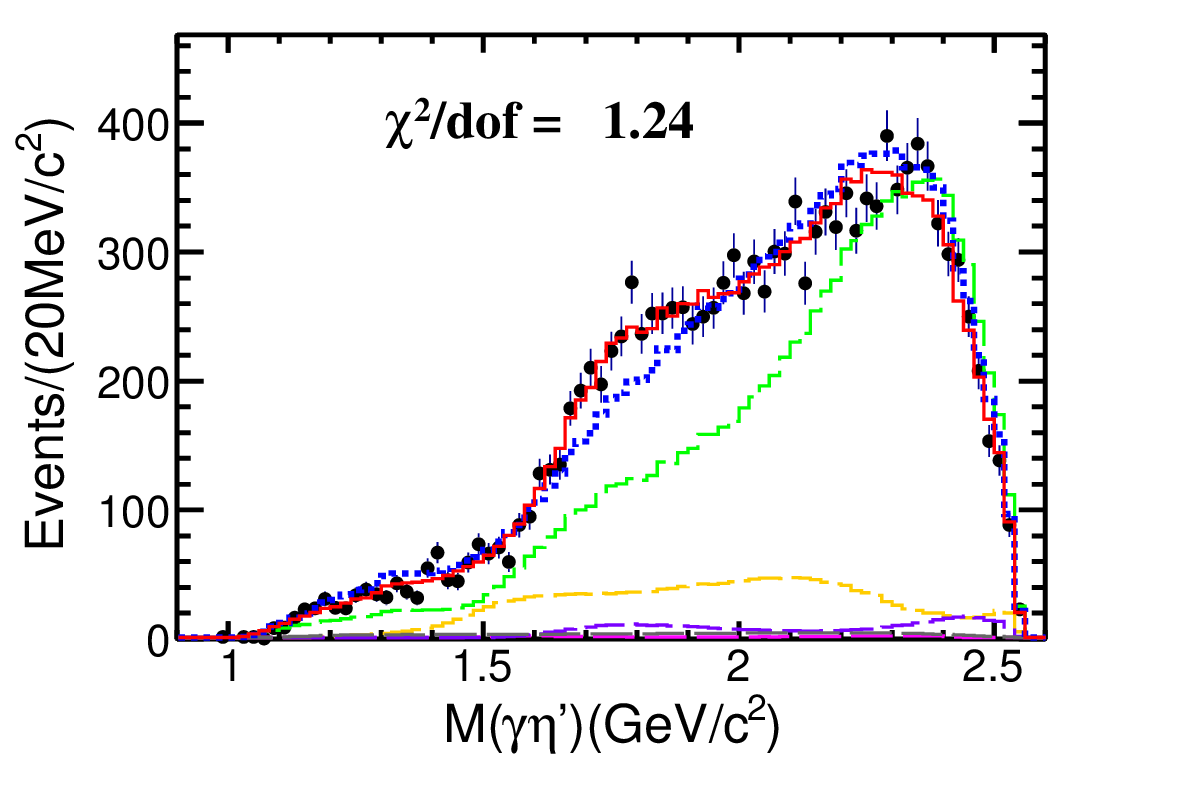}\put(-129,97){(c)}} \\ 
   \subfigure{
     \includegraphics[width=0.34\textwidth]{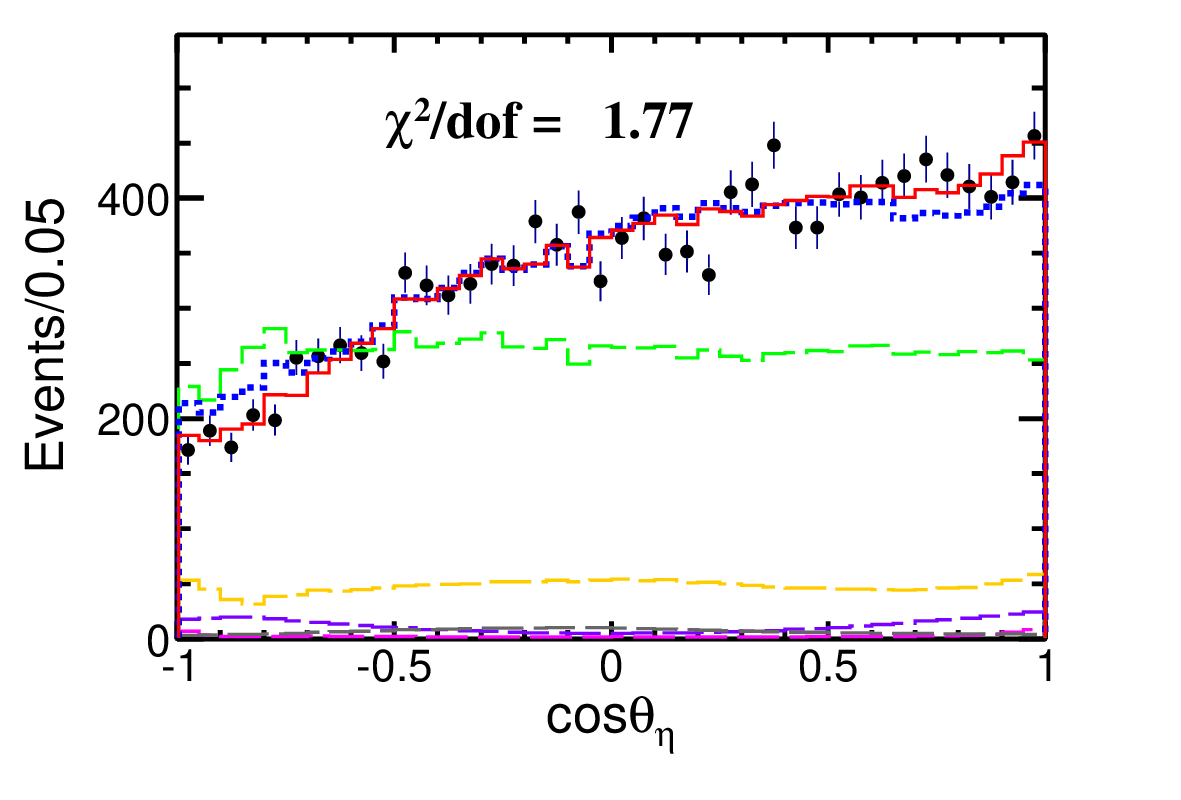}\put(-129,97){(d)}}
   \subfigure{
     \includegraphics[width=0.34\textwidth]{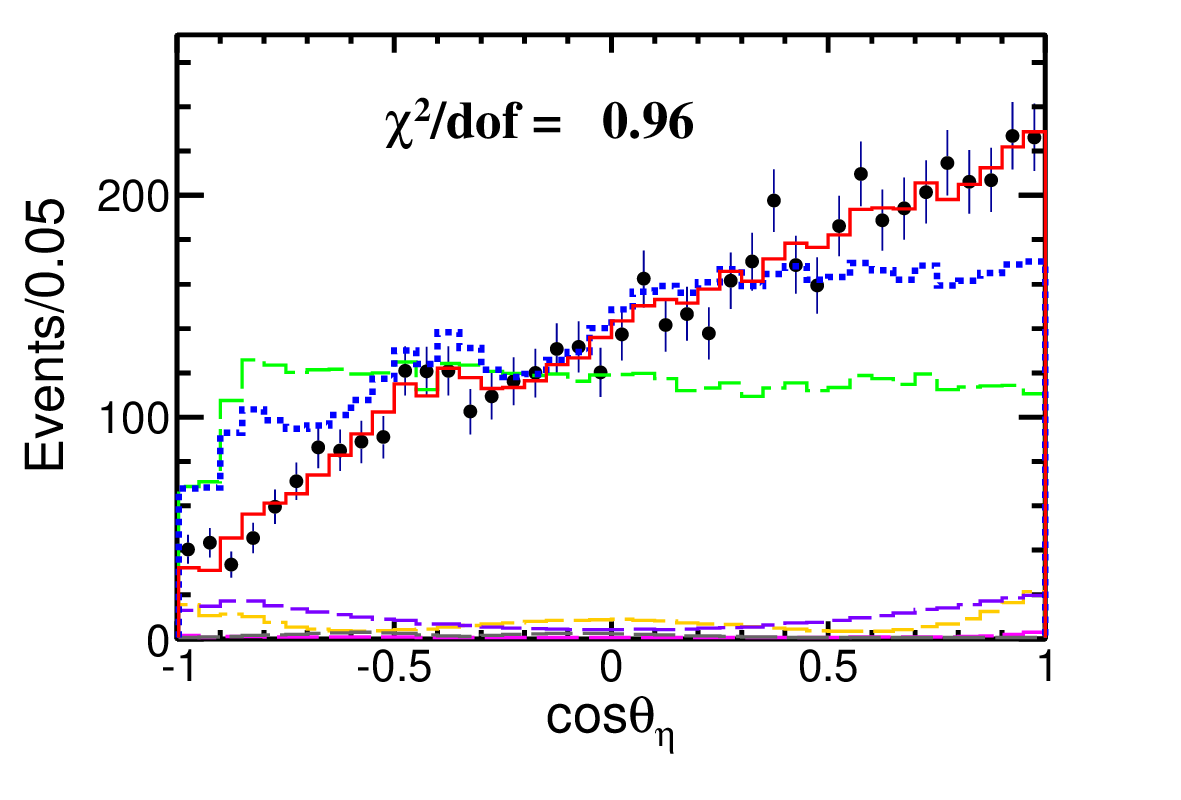}\put(-129,97){(e)}} \\ 

   \caption{Background-subtracted data (black points) and the PWA fit projections (lines) for (a),(b),(c) the invariant mass distributions of 
(a) $\eta\etap$, 
(b) $\gamma\eta$,
and (c) $\gamma\eta'$, and (d),(e) the distribution of cos$\theta_{\eta}$,  
where $\theta_{\eta}$ is the angle of the $\eta$ momentum in the $\eta\etap$ (Jocob and Wick) helicity frame for (d) all $\eta\etap$ masses and (e) $\eta\etap$ masses between 1.7 and 2.0~GeV/$c^2$.
The red lines are the total fit projections from the baseline PWA.  The blue lines are the total fit projections from a fit excluding the $\eta_1$ component. The dashed lines for the $1^{-+}, 0^{++}, 2^{++}, 4^{++}$, and $1^{+-}$ contributions are the coherent sums of amplitudes for each $J^{PC}$.
Note that the process $\jpsi\rightarrow\phi\etap,\phi\rightarrow\gamma\eta$ is rejected, 
which leads to the depletion of events around 1.02 GeV/$c^2$ in M$(\gamma\eta)$.
}
  \label{PWA fit plot}
\end{figure*}

\begin{figure*}[htbp]
   \centering
   
     \subfigure{
          \includegraphics[width=0.32\textwidth]{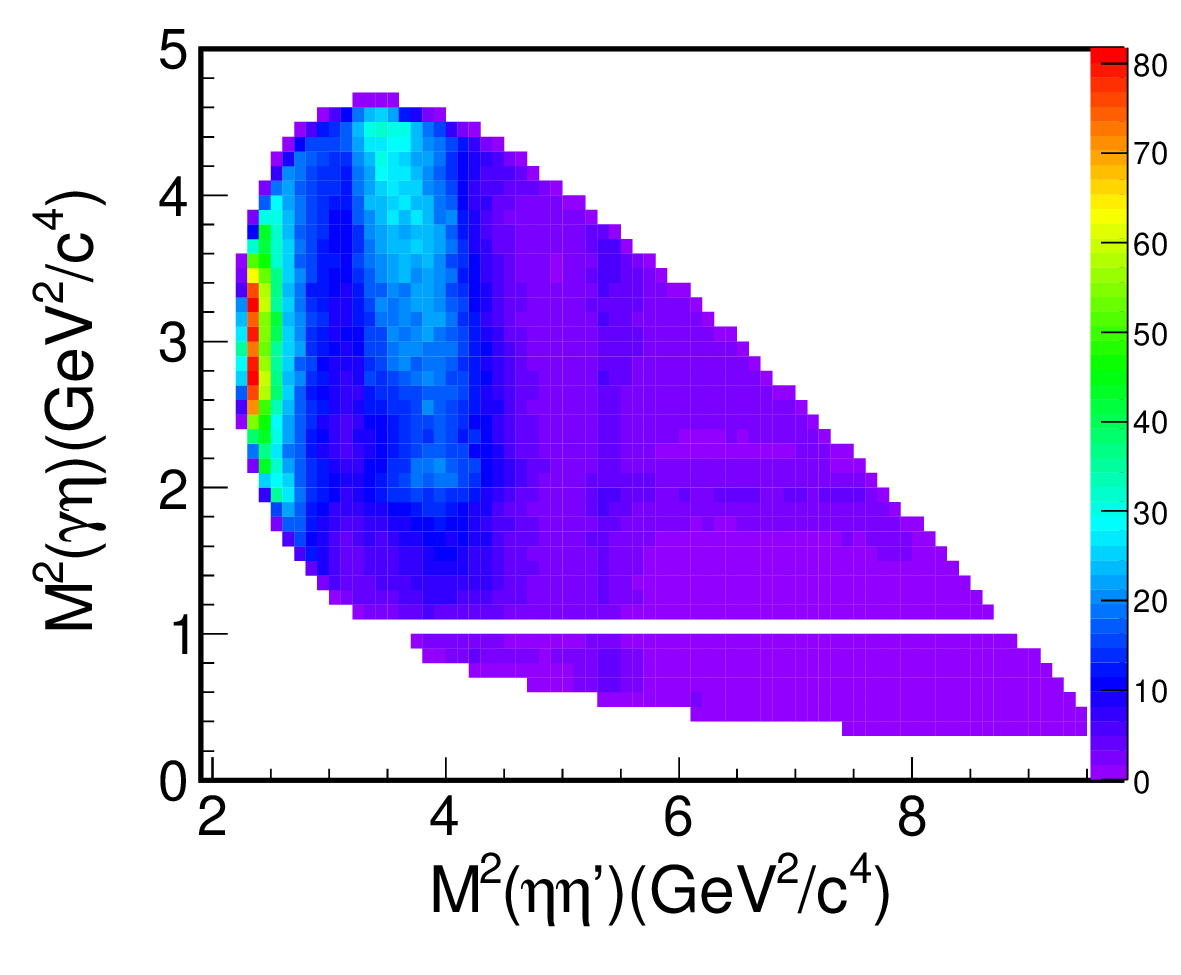}\put(-45,120){(a)}}
         \subfigure{
     \includegraphics[width=0.32\textwidth]{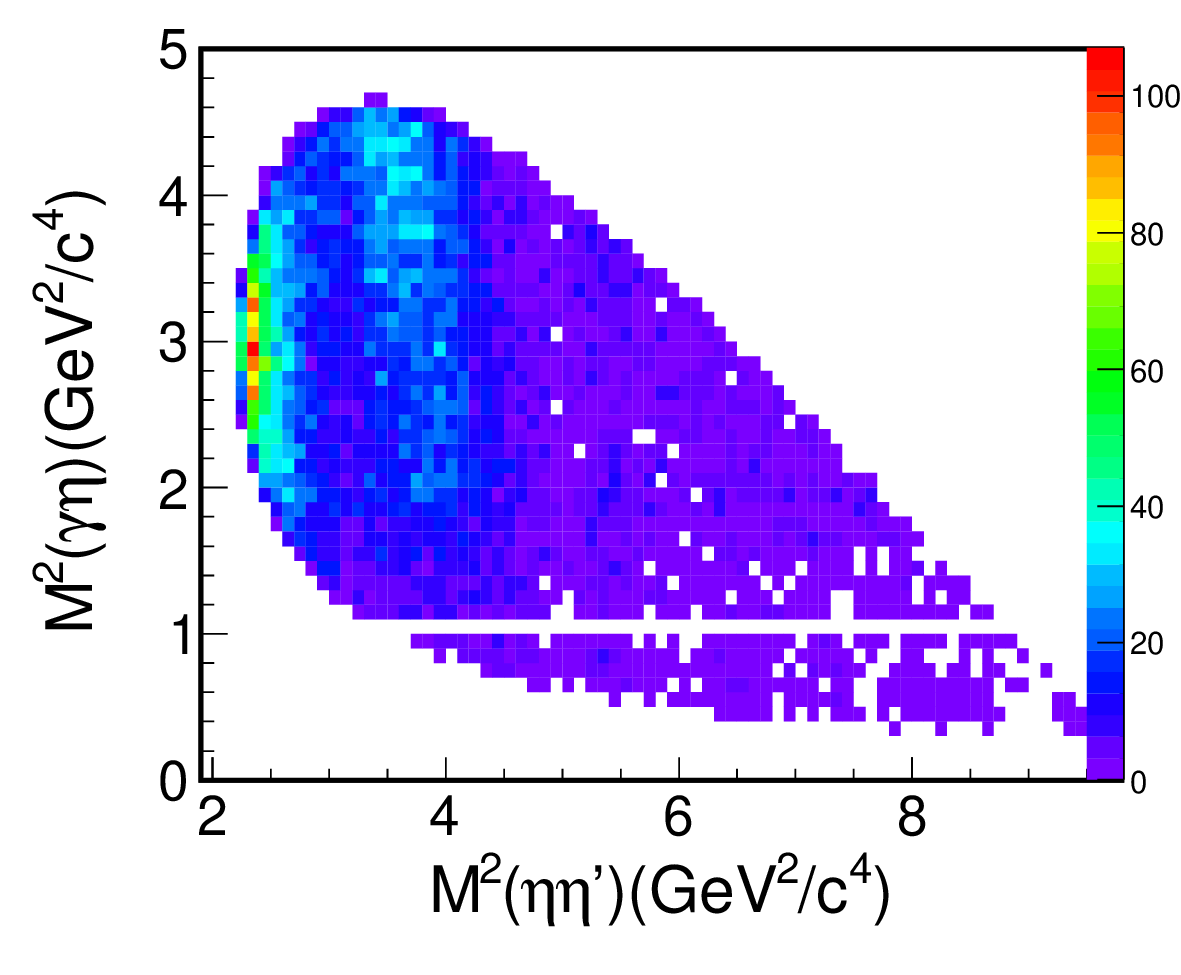}\put(-45,120){(b)}}
          \subfigure{
     \includegraphics[width=0.32\textwidth]{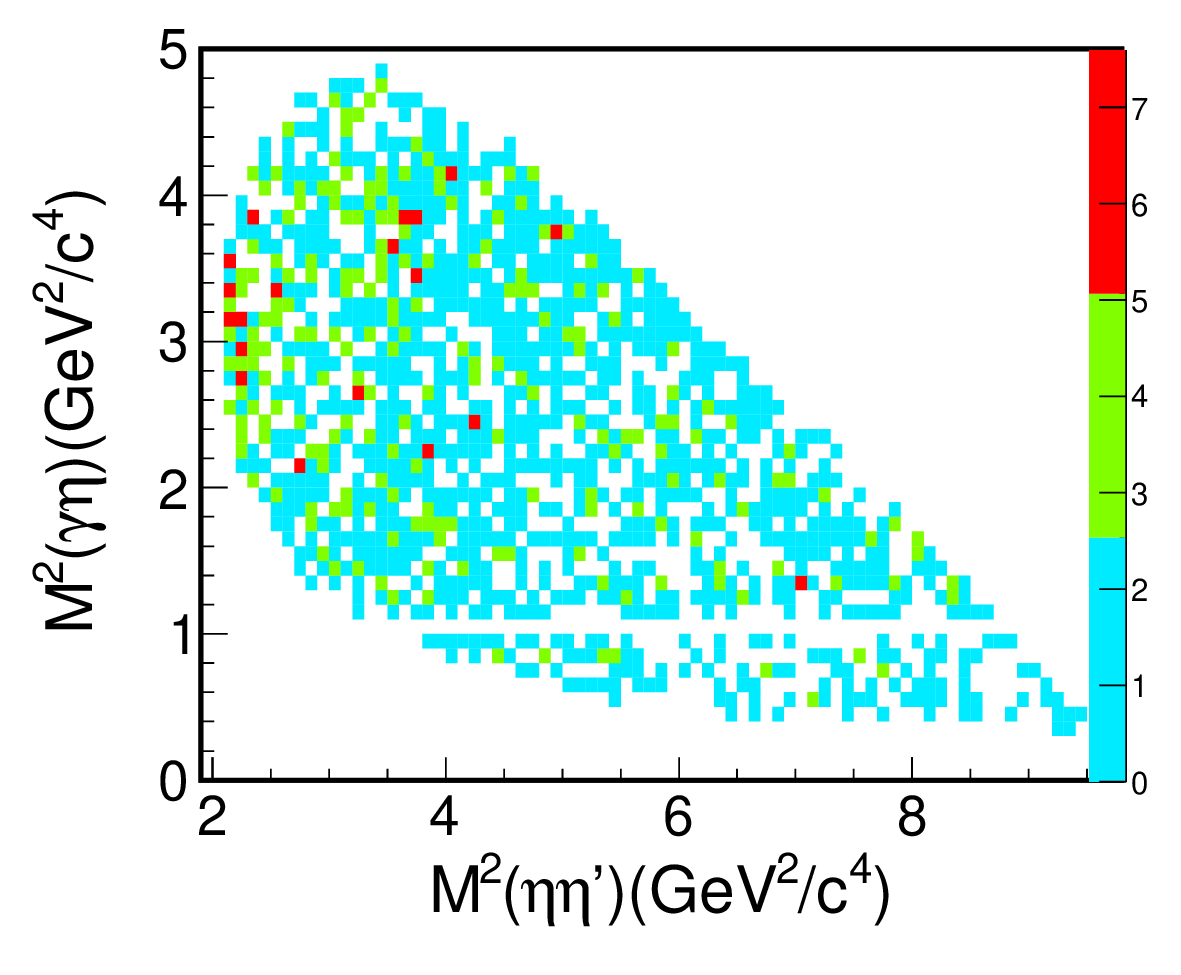}\put(-45,120){(c)}}\\

      \caption{Dalitz plots for  (a) the baseline PWA, (b) the selected data, and (c) background estimated from the $\etap$ sideband.
      Note that the process $\jpsi\rightarrow\phi\etap,\phi\rightarrow\gamma\eta$ is rejected, 
which leads to the depletion of events in a
band around 1 GeV$^2$/$c^4$ in M$^2(\gamma\eta)$.
      }
  \label{DalitzPlot}
  \end{figure*}

Figure~\ref{PWA fit plot} shows the  invariant mass distributions of $M(\eta\etap)$, $M(\gamma\eta)$, and $M(\gamma\etap)$ for the data (with background subtracted) and the PWA fit projections.  
Figure~\ref{PWA fit plot} also shows the cos$\theta_{\eta}$ distribution, where $\theta_{\eta}$ is the angle of the $\eta$ momentum in the $\eta\etap$ (Jocob and Wick) helicity frame~\cite{Jacob:1959at}.  This angle carries information about the spin of the particle decaying to $\eta\etap$.
Figure~\ref{DalitzPlot} shows the Dalitz plots for the PWA fit projection, the selected data, and the background estimated from the $\etap$ sideband.

\begin{table}[htbp]
\linespread{1.5}
\centering
\begin{small}
\caption{The masses, widths, $\BR(J/\psi\rightarrow \gamma X \rightarrow \gamma \eta\etap)$ or $\BR(J/\psi\rightarrow \etap h_1 \rightarrow \gamma \eta\etap)$ (B.F.), and statistical significances~(Sig.) 
for each component in the baseline set of amplitudes. The first uncertainties are statistical, and the second are systematic.  }\label{Summary of all}
\begin{tabular}{ccccc}
\hline
\hline
Resonance     &$M$ (MeV/$c^{2}$) &$\Gamma$ (MeV) &B.F.($\times$10$^{-5}$)  &Sig.  \\

\hline
$f_{0}(1500)$  &     1506        &     112                                  &1.81$\pm0.11_{-0.13}^{+0.19}$   &$\textgreater$30$\sigma$             \\
\hline

$f_{0}(1810)$  &   1795           &         95                  &0.11$\pm$$0.01_{-0.03}^{+0.04}$    &11.1$\sigma$            \\

\hline

$f_{0}(2020)$  &    2010$\pm$6$_{-4}^{+6}$         &           203$\pm$9$_{-11}^{+13}$                  &     2.28$\pm$0.12$_{-0.20}^{+0.29}$          &24.6$\sigma$     \\
\hline
$f_{0}(2330)$  &    2312$\pm$7$_{-3}^{+7}$         &               65$\pm$10$_{-12}^{+3}$           &     0.10$\pm$0.02$_{-0.02}^{+0.01}$        &13.2$\sigma$      \\
\hline
$\eta_{1}(1855)$  &     1855$\pm$9$_{-1}^{+6}$        &             188$\pm$18$_{-8}^{+3}$                       &       0.27$\pm$$0.04 _{-0.04}^{+0.02}$     &21.4$\sigma$       \\

\hline
$f_{2}(1565)$  &     1542                                &              122                             &           0.32$\pm$0.05$_{-0.02}^{+0.12}$     &8.7$\sigma$      \\
\hline
$f_{2}(2010)$  &     2062$\pm$6$_{-7}^{+10}$       &               165$\pm$17$_{-5}^{+10}$                       &       0.71$\pm$0.06$_{-0.06}^{+0.10}$         &13.4$\sigma$     \\

\hline
$f_{4}(2050)$  &      2018        &   237                                    &         0.06$\pm$0.01$_{-0.01}^{+0.03}$        &4.6$\sigma$   \\
\hline
$0^{++}$ PHSP  &     ...         &   ...                                    &         1.44$\pm$0.15$_{-0.20}^{+0.10}$         &15.7$\sigma$\\
\hline
$h_{1}(1415)$    &       1416       &    90                                   &       0.08$\pm$0.01$_{-0.02}^{+0.01}$          &10.2$\sigma$    \\
\hline
$h_{1}(1595)$    &        1584      &    384                                   &    0.16$\pm$0.02$_{-0.01}^{+0.03}$            &9.9$\sigma$     \\
\hline
\hline
\end{tabular}
\end{small}
\end{table}

Various checks are performed to validate the existence of the $\etamp$. 
The fits are carried out by assigning all other possible $J^{PC}$ ($J\leq4$) to the $\eta_{1}(1855)$, and the log-likelihoods are worse by at least 235 units ($\textgreater$30$\sigma$). 
To probe the significance of the BW phase motion, the BW parametrization of the $\etamp$ in the baseline PWA is replaced with an amplitude whose magnitude matches that of a BW function but with constant phase (independent of $s$). This alternative fit has a log-likelihood 43 units (9.2$\sigma$) worse than the baseline fit. 

\begin{figure*}[htbp]
\centering
 \subfigure{
 \label{Y00}
\includegraphics[width=0.38\textwidth]{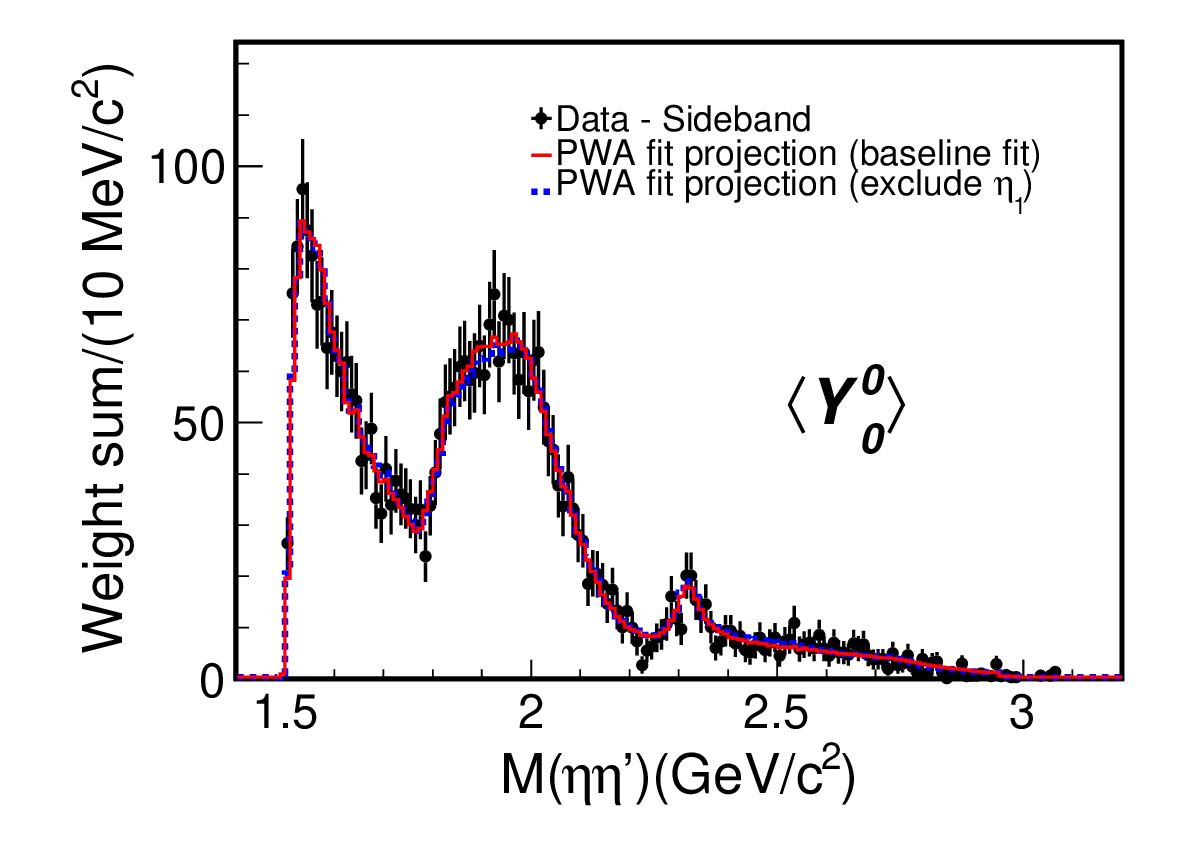}\put(-25,120){(a)}} 
\subfigure{
\label{Y01}
\includegraphics[width=0.38\textwidth]{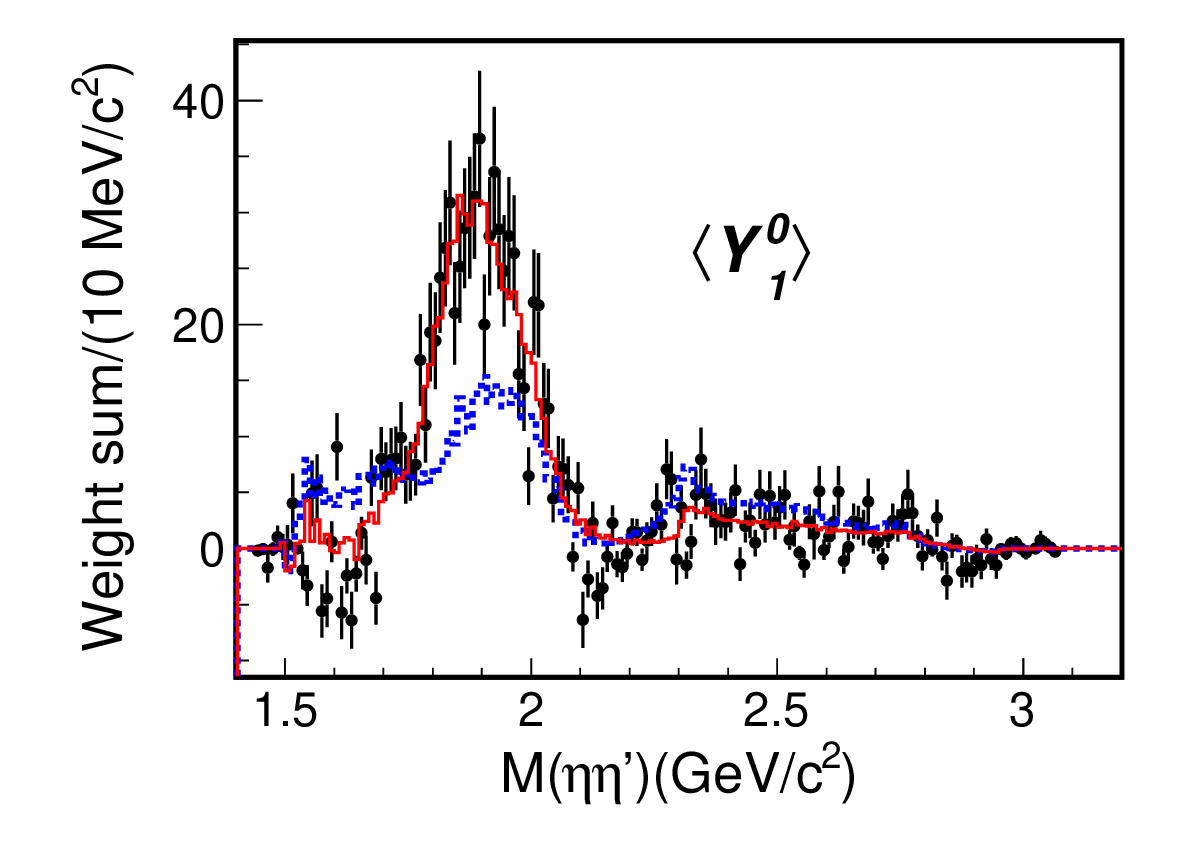}\put(-25,120){(b)}} 
\subfigure{
\label{Y02}
\includegraphics[width=0.38\textwidth]{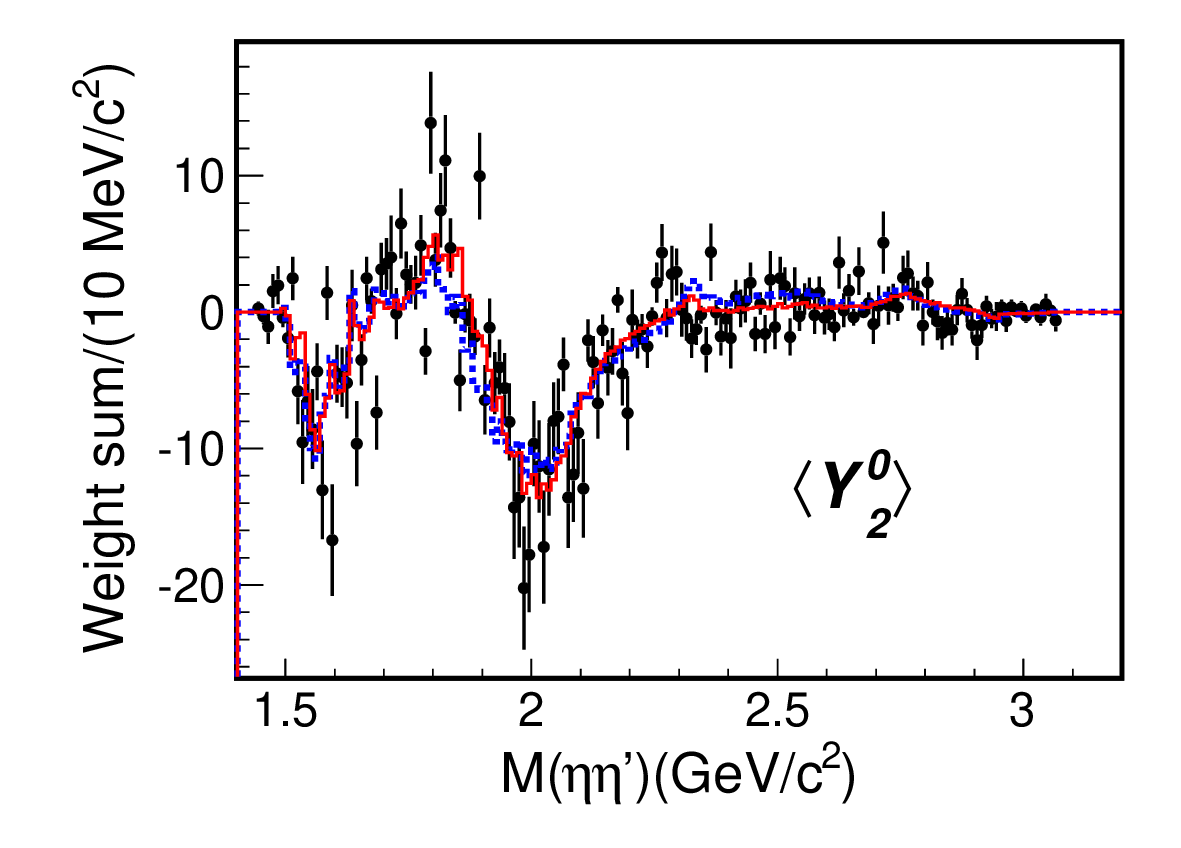}\put(-25,120){(c)}} 
\subfigure{
\label{Y04}
\includegraphics[width=0.38\textwidth]{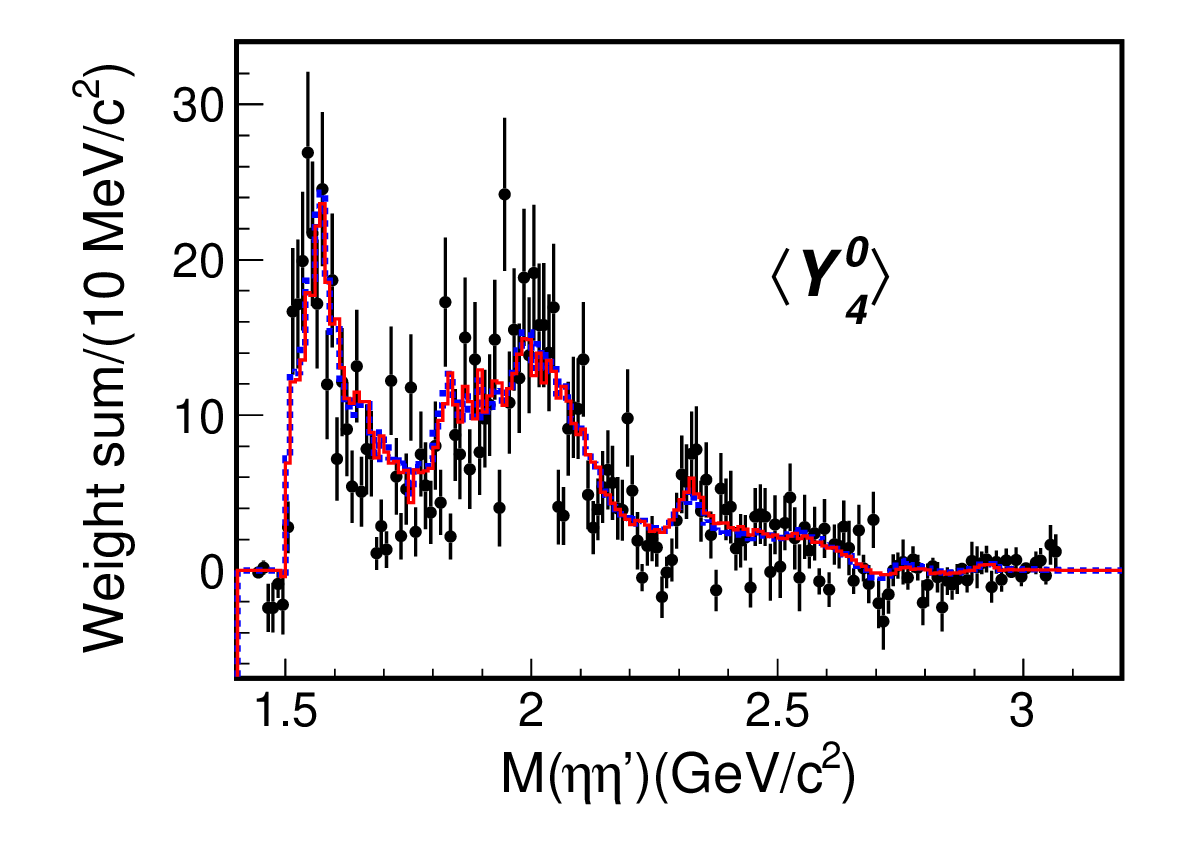}\put(-25,120){(d)}}

\caption{The distributions of the unnormalized moments 
$\langle Y^{0}_{L} \rangle$ ($L=$ 0, 1, 2, and 4) for $\jpsi\rightarrow\gamma\eta\etap$ as functions of the $\eta\etap$ mass. Black dots with error bars represent the background-subtracted data weighted with angular moments; the red solid lines represent the baseline fit projections; and the blue dotted lines represent the projections from a fit excluding the $\eta_1$ component.}
\label{Angular moments}
\end{figure*}

To visualize the agreement between the PWA fit results and data, angular moments as a function of $M(\eta\etap)$ can be calculated for data (with background subtracted) and the PWA model. 
For events within a given region of $M(\eta\etap)$, the cos$\theta_{\eta}$ distribution can be expressed as an expansion in terms of Legendre polynomials. 
The coefficients, 
which are called the unnormalized moments of the expansion, 
characterize the spin of the contributing $\eta\etap$ resonances. 
The moment for the $k$th bin of $M(\eta\etap)$ is 
\begin{eqnarray}\label{eqnY}
\langle Y^{0}_{l} \rangle \equiv \sum\limits_{i=1}^{N_{k}}W_{i}Y^{0}_{l}({\rm cos\theta}_{\eta}^{i}).
\end{eqnarray}
For data, $N_{k}$ is the number of observed events in the $k$th bin of $M(\eta\etap)$ and $W_{i}$ is a weight used to implement background subtraction. 
For the PWA model, $N_{k}$ is the number of events in a PHSP MC sample
and $W_{i}$ is the intensity for each event calculated in the PWA model. 
Neglecting $\eta\eta'$ amplitudes with spin greater than 2, and ignoring the effects of symmetrization and the presence of resonance contributions in the $\gamma\eta$ and $\gamma\eta'$ subsystems, 
the moments are related to the spin-0 ($S$), spin-1 ($P$) and spin-2 ($D$) amplitudes by~\cite{Costa:1980ji,PrivateCommunication}
\begin{widetext}
\begin{eqnarray}\label{eqnY00}
\sqrt{4\pi}\langle Y^{0}_{0} \rangle = S^{2}_{0} + P^{2}_{0} + P^{2}_{1} + D^{2}_{0} + D^{2}_{1} + D^{2}_{2},
\end{eqnarray}
\begin{eqnarray}\label{eqnY01}
\sqrt{4\pi}\langle Y^{0}_{1} \rangle = 2S_{0}P_{0} \cos\phi_{P_{0}} + \frac{2}{\sqrt{5}}( 2P_{0}D_{0}\cos(\phi_{P_{0}} - \phi_{D_{0}}) + \sqrt{3}P_{1}D_{1}\cos(\phi_{P_{1}} - \phi_{D_{1}} )) ,
\end{eqnarray}
\begin{eqnarray}\label{eqnY02}
\sqrt{4\pi}\langle Y^{0}_{2} \rangle = \frac{1}{7\sqrt{5}}(14P^{2}_{0} - 7P^{2}_{1} + 10D^{2}_{0} + 5D^{2}_{1} - 10D^{2}_{2}) + 2S_{0}D_{0}\cos\phi_{D_{0}}  ,
\end{eqnarray}
\begin{eqnarray}\label{eqnY04}
\sqrt{4\pi}\langle Y^{0}_{4} \rangle = \frac{1}{7}( 6D^{2}_{0} - 4D^{2}_{1} + D^{2}_{2}),
\end{eqnarray}
\end{widetext}
where $\phi_{P}$ and $\phi_{D}$ are the phases of the $P$ wave and $D$ wave relative to the $S$ wave.
Figure~\ref{Angular moments} shows the moments computed for the data and the PWA model, using Eq.~\ref{eqnY}, where good data and PWA consistency can be seen.  The need for the $\etamp$ $P$ wave component is apparent in the $\langle Y^{0}_{1} \rangle$ moment [Fig.~\ref{Angular moments}(b)].

For the branching fraction measurements, systematic uncertainties 
arising from 
the number of $\jpsi$ events, the pion tracking, photon detection, kinematic fit, mass resolution of the $\etap$, and the branching fractions of $ \etap\rightarrow\eta\pimp, \etap\rightarrow\gamma\pimp$, and $\eta\rightarrow\gamma\gamma$ have been estimated to be 4.8$\%$~\cite{PRD}.

Uncertainty associated with the PWA affects both the branching fraction measurements and the resonance parameters.
The sources of uncertainty include the 
background estimation, 
the resonance description, the resonance parameters,
and additional resonances. 
The statistical significance of the $\etamp$ is recalculated in each fit variation.

To estimate the uncertainty due to the background estimation,
alternative fits are performed using different background normalization factors and different $\etap$ sideband regions.
The statistical significance of the $\etamp$ is always above 21.1$\sigma$.
The changes in the branching fractions and resonance parameters are assigned as systematic uncertainties.

Uncertainty arising from the BW parametrization is estimated by replacing the constant width $\Gamma_{0}$ of the BW for the threshold state $f_{0}(1500)$ with a mass-dependent width as described in Ref.~\cite{PRD}. The significance of the $\etamp$ in this case is 21.8$\sigma$.

In the baseline fit, the resonance parameters of  the $f_{0}(1500)$, $f_{0}(1810)$, $f_{2}(1565)$, $f_{4}(2050)$, $h_{1}(1415)(\gamma\eta)$, and $h_{1}(1595)(\gamma\eta)$ are fixed to PDG~\cite{Zyla:2020zbs} average values. An alternative fit is performed where resonance parameters are allowed to vary within 1 standard deviation of the PDG values~\cite{Zyla:2020zbs}, and the changes in the results are taken as systematic uncertainties. 
The statistical significance of the $\etamp$ in this case is 20.6$\sigma$.

Uncertainties arising from possible additional resonances are estimated by adding the $f_0(1710)$, $f_2(2220)$, $f_4(2300)$, $h_1(1595)(\gamma\eta')$, and $\rho(1900)(\gamma\eta')$, which are the most significant additional resonances for each possible $J^{PC}$, into the baseline fit individually. 
The resulting changes in the measurements are assigned as systematic uncertainties. In all cases, the significance of the $\etamp$ remains larger than 19.0$\sigma$.

Assuming all of these sources are independent, the total systematic uncertainties are $^{+6}_{-1}$~MeV/$c^{2}$ and $^{+3}_{-8}$~MeV for the  mass and width of the $\etamp$, respectively.  For the branching fraction of the $\etamp$, the total relative systematic uncertainty is determined to be $^{+5.9}_{-13.1}$$\%$. Tables VII and VIII of Ref.~\cite{PRD} summarize the systematic uncertainties.

The ratios $\BR(f_0$$\rightarrow$$\eta\etap)$/$\BR(f_0$$\rightarrow$$\pi\pi)$ can be calculated with the branching fractions measured in this analysis and PDG~\cite{Zyla:2020zbs}. 
The ratio $\BR(f_0(1500)$$\rightarrow$$\eta\etap)$/$\BR(f_0(1500)$$\rightarrow$$\pi\pi)$ is determined to be (1.66$^{+0.42}_{-0.40}$)$\times$10$^{-1}$, where the error is the combined systematic and statistical uncertainties. 
In comparison, the upper limit on  $\BR(f_0(1710)$$\rightarrow$$\eta\etap)$/$\BR(f_0(1710)$$\rightarrow$$\pi\pi)$ at 90$\%$ confidence level is determined to be 2.87$\times$10$^{-3}$.
The suppressed decay rate of $f_0(1710)$ into $\eta\eta'$ is further discussed in Ref.~\cite{PRD}

In summary, a PWA of $J/\psi\rightarrow\gamma\eta\etap$ has been performed based on (10.09$\pm$0.04)$\times$10$^{9}$ $\jpsi$ events collected with the BESIII detector. An isoscalar state with exotic quantum numbers $J^{PC}=1^{-+}$, denoted as $\etamp$, has been observed for the first time. 
The statistical significance of the resonance hypothesis is estimated to be larger than 19$\sigma$. The product branching fraction  $\BR(J/\psi$$\rightarrow$$ \gamma\eta_1(1855))$$\BR(\eta_1(1855)\rightarrow$$\eta\eta')$ is measured to be (2.70$\pm  0.41 _{-0.35}^{+0.16}) \times$10$^{-6}$.
Its mass and width are measured to be (1855$\pm$9$_{-1}^{+6}$)~MeV/$c^{2}$ and (188$\pm$18$_{-8}^{+3}$)~MeV, respectively. 
The first uncertainties are statistical and the second are systematic.
The mass and width of the $\eta_1(1855)$ are consistent with LQCD calculations for the $1^{-+}$ hybrids~\cite{Dudek:2013yja}. 
The observation of the isoscalar $\eta_1(1855)$, combined with previous measurements of the isovector $\pi_1$ states, provides critical information about the $1^{-+}$ hybrid nonet.
Further studies with more production mechanisms and decay modes will help clarify the nature of the $\eta_{1}(1855)$.

The BESIII collaboration thanks the staff of BEPCII and the IHEP computing center for their strong support. This work is supported in part by National Key R$\&$D Program of China under Contracts Nos. 2020YFA0406300, 2020YFA0406400; National Natural Science Foundation of China (NSFC) under Contracts Nos. 11625523, 11635010, 11675183, 11735014, 11822506, 11835012, 11922511, 11935015, 11935016, 11935018, 11961141012, 12022510, 12025502, 12035009, 12035013, 12061131003; the Chinese Academy of Sciences (CAS) Large-Scale Scientific Facility Program; the CAS Center for Excellence in Particle Physics (CCEPP); Joint Large-Scale Scientific Facility Funds of the NSFC and CAS under Contracts Nos. U1732263, U1832207; CAS Key Research Program of Frontier Sciences under Contract No. QYZDJ-SSW-SLH040; 100 Talents Program of CAS; INPAC and Shanghai Key Laboratory for Particle Physics and Cosmology; ERC under Contract No. 758462; European Union Horizon 2020 research and innovation programme under Contract No. Marie Sklodowska-Curie grant agreement No 894790; German Research Foundation DFG under Contracts Nos. 443159800, Collaborative Research Center CRC 1044, GRK 214; Istituto Nazionale di Fisica Nucleare, Italy; Ministry of Development of Turkey under Contract No. DPT2006K-120470; National Science and Technology fund; Olle Engkvist Foundation under Contract No. 200-0605; STFC (United Kingdom); The Knut and Alice Wallenberg Foundation (Sweden) under Contract No. 2016.0157; The Royal Society, UK under Contracts Nos. DH140054, DH160214; The Swedish Research Council; U. S. Department of Energy under Contracts Nos. DE-FG02-05ER41374, DE-SC-0012069.
\nocite{*}
\bibliographystyle{apsrev4-1}
\bibliography{gammaEtaEtap_PRL}

\end{document}